# Dichotomy of the transport coefficients of correlated electron liquids in SrTiO$_3$

**Tyler A. Cain, Evgeny Mikheev, Clayton A. Jackson, and Susanne Stemmer[a)]**

Materials Department, University of California, Santa Barbara, CA 93106-5050, USA

[a)] Corresponding author: stemmer@mrl.ucsb.edu

**Abstract**

We discuss the Seebeck coefficient and the Hall mobility of electrons confined in narrow $SrTiO_3$ quantum wells as a function of the three-dimensional carrier density and temperature. The quantum wells contain a fixed sheet carrier density of $\sim 7 \times 10^{14}$ cm$^{-2}$ and their thickness is varied. At high temperatures, both properties exhibit apparent Fermi liquid behavior. In particular, the Seebeck coefficient increases nearly linearly with temperature ($T$) when phonon drag contributions are minimized, while the mobility decreases proportional to $T^2$. Furthermore, the Seebeck coefficient scales inversely with the Fermi energy (decreasing quantum well thickness). In contrast, the transport scattering rate is independent of the Fermi energy, which is inconsistent with a Fermi liquid. At low temperatures, the Seebeck coefficient deviates from the linear temperature dependence for those electron liquids that exhibit a correlation-induced pseudogap, indicating a change in the energy dependence of the scattering rate. The implications for describing transport in strongly correlated materials are discussed.

The transport properties of strongly correlated materials in the normal (non-superconducting) state reflect the complexity of interactions of the charge carriers in these materials. Of particular interest are deviations from Fermi liquid behavior, because these can be signatures of strong electron correlation physics [1, 2]. SrTiO$_3$ is emerging as a promising system for such studies. For example, the resistance ($R$) of doped SrTiO$_3$ scales with $T^2$ over a temperature ($T$) range that expands with doping [3-7]. The quadratic temperature dependence is commonly associated with electron-electron scattering in a Fermi liquid. For highly doped, bulk SrTiO$_3$ and high-density, two-dimensional electron liquids in SrTiO$_3$, $R \sim T^2$ to room temperature [5, 8]. Several other observations are, however, inconsistent with a Fermi liquid [7]. For example, $R \sim T^2$ even at low carrier densities, where it is difficult to relax the momentum via Umklapp scattering [6], and at elevated temperatures, where the electron system may be non-degenerate ($E_F < k_B T$, where $E_F$ is the Fermi energy and $k_B$ the Boltzmann constant) [5, 7, 9]. Furthermore, the scattering rate is found to be independent of the carrier density and the Fermi energy ($E_F$) [7]. This is in contrast to Fermi liquid theory, where the scattering rate is proportional to $(k_B T)^2/E_F$ [10]. If the peculiar behavior of the transport scattering rate would be a feature only of SrTiO$_3$, one may consider this to be merely a curiosity. $R \sim T^2$ is ubiquitous in strongly correlated materials, including the parent (overdoped) phase of high-temperature superconductors, where it is often thought to signify a Fermi liquid. Strikingly, the density-independent scattering rate in the $T^2$ regime is not unique to SrTiO$_3$; it has also been observed in other correlated materials [11-13]. The irrelevance of $E_F$ in the scattering rate is, of course, explicit in case of the famous $T$-linear scattering rate of the cuprate superconductors, where $T$ becomes the only energy scale [14]. The ubiquity of poorly understood, carrier-density independent scattering rates in very different materials makes doped SrTiO$_3$ an intriguing, relatively simple, model system for developing an understanding of the transport properties of correlated electron liquids.

An obvious next step is to investigate other transport coefficients. The Seebeck coefficient is the ratio of the thermoelectric and electrical conductivities [15]. In a Fermi liquid, it becomes independent of the scattering rate, *if* all carriers have the same mean free path, independent of their energy. Non-diffusive (phonon drag) contributions dominate the Seebeck coefficient in doped $SrTiO_3$ below room temperature over a wide range of carrier densities [16, 17], which makes quantitative interpretation challenging. Here, we study the Seebeck coefficient of electron liquids embedded in $SrTiO_3$ quantum wells with sheet carrier densities on order of $7\times10^{14}$ $cm^{-2}$. As will be shown here, these quantum wells are within the degenerately doped regime and phonon drag becomes negligible in sufficiently narrow quantum wells. This allows us to compare the temperature and carrier density dependence of the Seebeck coefficient and of the Hall mobility with those expected for a Fermi liquid. Discrepancies arise in the transport scattering rate and in the Seebeck coefficient at low temperature. While the latter can be explained with an energy-dependent scattering rate that changes with temperature and that is associated with a pseudogap, no simple explanation exists for the former.

$SrTiO_3$ quantum wells with a sheet carrier density $\sim 7\times10^{14}$ $cm^{-2}$, which is fixed by the interfacial polar discontinuity [18, 19], were embedded either in ferrimagnetic, insulating $GdTiO_3$ or in antiferromagnetic, insulating $SmTiO_3$. Their structure, magnetic, and transport properties have been characterized extensively, as described elsewhere [8, 18, 20-22]. Here, the thicknesses were 10 nm for the $SmTiO_3$ layers, 4 nm for the $GdTiO_3$ layers. We vary the $SrTiO_3$ quantum well thickness, which we specify in number of SrO layers per quantum well. Because the sheet carrier density is fixed, reducing the thickness of the quantum well increases the three-dimensional carrier density and $E_F$. For temperature-dependent electrical measurements, contacts were deposited by electron beam evaporation and were 40 nm Ti/400 nm Au, where the Ti contacts the quantum well. All measurements were carried out in a Physical Properties Measurement System (Quantum Design PPMS). Care was taken to correct for artifacts in the Seebeck coefficients measured with the PPMS Thermal Transport Option, as described in ref. [23].

Figure 1 shows the temperature coefficient, $\alpha$, of the inverse of the Hall mobility, $\mu^{-1}$, as a function of the SrTiO$_3$ quantum well thickness:

$$\mu^{-1} = \mu_0^{-1} + \alpha T^n, \qquad (1)$$

where $\mu_0$ is the residual due to disorder (defect) scattering. Here, $n \approx 2$ for all quantum wells [8]. In contrast, the longitudinal resistance shows anomalous $n \approx 1.6$ for SmTiO$_3$ thicknesses close to 5 SrO layers for quantum wells in SmTiO$_3$, which is not observed for quantum wells in GdTiO$_3$ [8, 24]. (The discrepancy of the power law exponents, $n$, in the temperature dependencies of $R$ and $\mu^{-1}$ is known as lifetime separation [25, 26] and we have discussed it elsewhere [8]). As can be seen from Fig. 1, except for the thinnest quantum wells, which show mass enhancement due to the proximity to a metal-insulator transition [18, 20], $\alpha$ is independent of thickness. Because reducing the thickness increases the three-dimensional carrier density, this implies that $\alpha$ is independent of the three-dimensional carrier density and $E_F$.

The second term on the right hand side of Eq. (1) can also be expressed in terms of a characteristic scattering energy, $E$ [7, 10]:

$$\alpha T^n = \frac{m^*}{e} B \frac{(k_B T)^n}{(\hbar E)^{n-1}}, \qquad (2)$$

where $m^*$ is the effective mass, $e$ is the electron charge, $k_B$ the Boltzmann constant. In Fermi liquid theory, $E = E_F$ and $n = 2$ [10]. $B$ is a dimensionless scattering amplitude usually not far from unity [10, 14]. A carrier-independent scattering rate, as observed here, implies that $E = constant \neq E_F$. The same is true for the temperature coefficient of the longitudinal resistance, $R$ (see Supplementary Information of ref. [8]). As mentioned in the introduction, the scattering rare therefore appears inconsistent with electron-electron scattering in a Fermi liquid, even though $n = 2$.

Figure 2 shows the magnitude of the Seebeck coefficient, $|S|$ ($S < 0$ for all quantum wells) and $|S|/T$ for quantum wells of different thickness. The thicker quantum wells (> 6 SrO layers) appear to have non-diffusive contributions to $|S|$, similar to bulk $SrTiO_3$. These enhance $|S|$ in a way not described by simple models [27] and we do not consider them further here. Above ~150 K, $|S|$ varies linearly with $T$ ($|S|/T = constant$) for the thinner quantum wells (See dashed lines in Fig. 2. A small phonon drag contribution is still visible for the two thickest quantum wells for which dashed lines are shown). Furthermore, $|S|/T$ decreases with quantum well thickness, i.e., as the three-dimensional carrier density and $E_F$ are raised by confinement. Below ~150 K, $|S|$ deviates from $T$-linear behavior in the quantum wells in $SmTiO_3$. In contrast, the quantum wells in $GdTiO_3$ remain Fermi-liquid-like ($|S|/T = constant$), with the exception of the 5 SrO quantum well [different 5 SrO samples were measured to confirm that this was not an outlier].

We first discuss the high temperature (> 150 K) behavior. A constant $|S|/T$ suggests a Fermi liquid, for which $S$ is given as [15]:

$$S = \frac{\pi^2}{3} \frac{k_B}{e} \frac{T}{T_F} (1 + \lambda), \tag{3}$$

where $T_F$ the Fermi temperature ($T_F = E_F/k_B$) and $\lambda$ is the energy ($E$) dependence of the mean free path, $l$, and is given as [15]:

$$\lambda = \frac{\partial \ln(l)}{\partial \ln(E)}\bigg|_{E=E_F}. \qquad (4)$$

Figure 2 shows that with decreasing quantum well thickness, $|S|/T$ decreases. Thus, unlike $\alpha$, $S$ scales inversely with $E_F$, as expected from Eq. (3). Figure 3 shows the extracted values for $E_F$ using the data from Fig. 2, Eq. (3), and setting $\lambda = 0$. While there are significant errors due to rather simple assumptions (single, parabolic band, constant mean free path), these should only affect the quantitative values and not the qualitative trend, namely that as the quantum well thickness decreases, the Seebeck coefficient shows that $E_F$ increases. We note that the quantitative change in $E_F$ with quantum well thickness is quite complicated due to a combination of factors, including the different orbital character of occupied subbands [28-30] as a function of thickness, in particular as the two electron liquids start to overlap significantly below 5 SrO layers [8].

The results illustrate why these electron systems remain so difficult to understand. Three observations, namely $|S| \sim T$ and $R, \mu^{-1} \sim T^2$, and the systematic change of $S$ with $E_F$, all suggest a Fermi liquid. In a Fermi liquid, the constant $|S|/T$ and the $T^2$-dependence of the resistance have the same origin: they are both a result of the Pauli exclusion principle, i.e., only electrons within $k_BT$ of the Fermi surface contribute to the thermoelectric and electrical conductivities, respectively. Nevertheless, the low energy excitations are controlled by an energy scale that is lower than $E_F$. From Eq. (2), taking $\alpha \sim 0.01$ Vsm$^{-2}$K$^{-2}$ (Fig. 1) and for $m^* \sim m_e$ ($m_e$ is the free electron mass), $E$ is about 6 meV. This is smaller than $E_F$, which is 120 - 300 meV (Fig. 3), even considering that these are order-of-magnitude estimates only. Moreover, the energy scale $E$ in

Eq. (2) is similar in the quantum wells and in bulk, lightly-doped SrTiO$_3$ [7]. In a previous study we found that in the low-temperature mobility of SrTiO$_3$ is highly sensitive to strain whereas in the $T^2$ regime it is not [31]. Combined, all of these results show that when the electronic structure is changed by a variety of tuning parameters (doping, quantum well thickness, strain), this is not (or barely) reflected in the transport scattering rate in the regime where $R$ and $\mu^{-1}$ follow a power law. The origin of the $T^2$ behavior and what determines the transport scattering rate remain thus very poorly understood. In some sense, the challenge posed by the electrical transport is similar to the ultrashort transport lifetimes in metals that exhibit $T$-linear resistance, which includes correlated materials as well as simple metals in the electron-phonon scattering limited regime. In these materials, the electron lifetime $\tau$ is determined by the uncertainty relation, $\tau \Delta E \sim \hbar$, with $\Delta E$ being the thermal energy distribution at the Fermi surface, $k_B T$, independently of what must be very different electronic structures [14, 32]. A common feature appears to be the very high scattering rates (small $E$ in the present study). Future studies should address what sets the energy scale $E$, as it likely contains the interaction physics. A quantitative understanding of the transport scattering rate will most likely require consideration of the phonon drag. As has been shown long ago for simple metals [33], this reduces electron-phonon scattering contributions to the resistance, making electron-electron scattering more important.

In contrast to the transport scattering rate, as discussed above, the Seebeck coefficient behaves, at least qualitatively, as expected, reflecting the change in $E_F$. Unlike the scattering rate, $S$ also reflects whether the electron gas is degenerate or not: for example, the carrier density dependence of the room temperature Seebeck coefficient of bulk doped SrTiO$_3$ indicates that it is non-degenerate [9], while in the quantum wells it behaves like a degenerate electron liquid (we note that in both cases the transport is metallic, $\delta R/\delta T > 0$, due to the large Bohr radius).

Finally, we address the low temperature (< 150 K) Seebeck coefficient. While $|S|/T$ remains constant in the quantum wells in GdTiO$_3$ (with the exception of the 5 SrO quantum well), it deviates from $T$-linearity for the quantum wells in SmTiO$_3$. A variation in $\lambda$ (Eq. (4)), the energy dependence of the mean free path, with temperature is a possible explanation. As can be seen from Fig. 4, which shows the differences between $|S|/T$ at 20 K and at 300 K, the deviation from the high-temperature $|S|/T$ value becomes significant at thicknesses below 5 SrO layers. In quantum wells in SmTiO$_3$, this coincides with the previously established critical thickness of 5 SrO layers for the quantum critical or Lifshitz point [8], the crossover to a non-Fermi liquid exponent $n \approx 1.6$ in the longitudinal scattering rate [8, 24] and the emergence of pseudogaps in the tunneling density of states [34, 35]. The pseudogap evolves similar to that in other correlated materials, including exhibiting a pile-up of states just outside the gap at the lowest temperatures [34]. Pseudogaps of this type are often associated with fluctuations of an order parameter and emergence of coherence at low temperatures [35, 36]. The Seebeck results therefore suggest that these are also associated with a temperature-dependent change in $\lambda$. Furthermore, electron correlations can result in $\lambda < 0$ [37]. Conversely, the loss of quasiparticle density of states in the quantum wells in GdTiO$_3$, which is not accompanied by coherence peaks [34], does not change $|S|/T$, except for the sample with 5 SrO layers, indicating that the energy dependence of the scattering rate does not change as a function of temperature. Thus, the different pseudogap behaviors, which are likely due to the different magnetic or orbital fluctuations/order in the two types of quantum wells, due to the coupling to the different barrier layers [34], are reflected in the Seebeck coefficient. While a complete understanding is still lacking, the experimental results for the quantum wells in SmTiO$_3$ show that correlation-induced pseudogaps, a longitudinal resistivity following a power law with $n < 2$, and deviations from a $T$-linear Seebeck coefficient due to a asymmetry in the energy dependence of the scattering rate, are all correlated. This points to the same microscopic origin, namely a specific type of fluctuation/order parameter that is closely coupled with antiferromagnetism and that may be relevant to other correlated materials that exhibit these non-Fermi liquid characteristics.

**Acknowledgements**

We gratefully acknowledge helpful discussions with Jim Allen, Oded Rabin, Jane Cornett, and Dmitrii Maslov. The work was supported by the U.S. Army Research Office (grant no. W911NF-14-1-0379) and made use of Central Facilities supported by the MRSEC Program of the U.S. National Science Foundation under Award No. DMR 1121053.

**Figure Captions**

**Figure 1:** Temperature coefficient of the inverse of the Hall mobility as a function of quantum well thickness (specified in terms of the number of SrO layers contained in the quantum well). For quantum wells in GdTiO$_3$, the fits were carried out between 2 – 300 K and for quantum wells in SmTiO$_3$ between 50 – 300 K (see ref. [8] for details).

**Figure 2:** Magnitude of the Seebeck coefficient (left) and $|S|/T$ (right) as a function of temperature and quantum well thickness (specified in number of SrO layers, see legend). (a,b) Quantum wells in SmTiO$_3$. (c,d) Quantum wells in GdTiO$_3$. The dashed lines are fits to the *T*-linear behavior in the appropriate temperature ranges and drawn over the entire temperature range to show any deviations from *T*-linear behavior.

**Figure 3:** The Fermi level, extracted from the measured Seebeck coefficient using Eq. 3, as a function of quantum well thickness.

**Figure 4:** Relative change of $|S|/T$ from 300 to 20 K as a function of quantum well thickness. The red dashed line indicates the 5 SrO layer thickness.

**Figure 1**

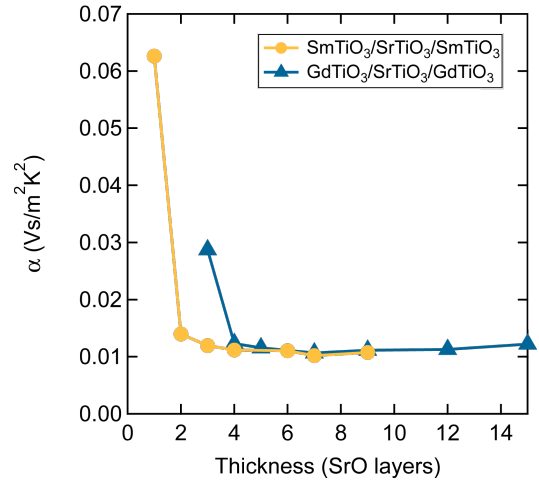

**Figure 2**

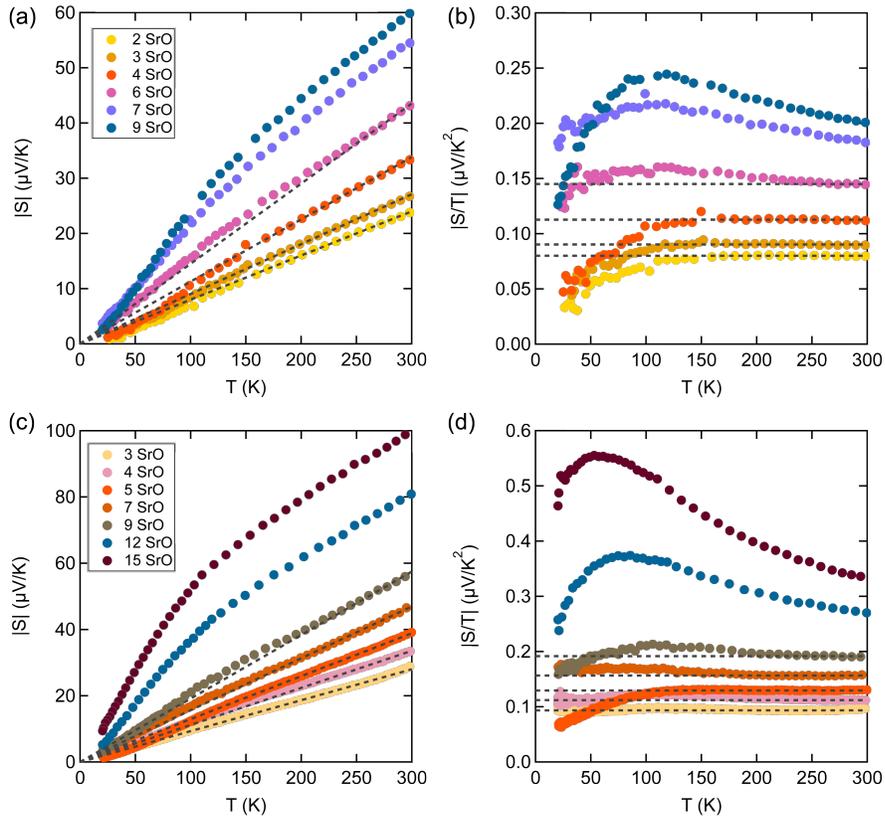

**Figure 3**

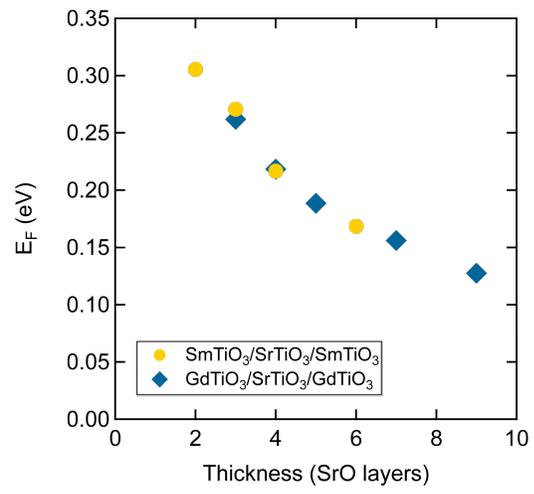

**Figure 4**

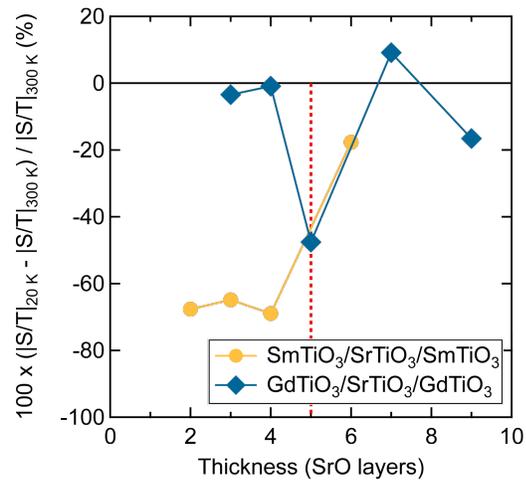